\input{aipcheck}

\documentclass[final
  ]
  {aipproc}

\layoutstyle{6x9}

\begin{document}
\newcommand{\nh}{$N_{\mathrm{H}}$}
\newcommand{\rxte}{\textit{RXTE}}
\newcommand{\inte}{\textit{INTEGRAL}}

\title{Ejection of the corona at State transitions: a common behaviour in microquasars?}

\classification{97.80.Jp}
\keywords      {accretion, accretion disks -- X-rays:binaries -- X-rays: individuals : XTE J1550-564, GRS 1915+105, GX 339-4, H1743-322}

\author{L. Prat}{
  address={DSM/IRFU/Service d'Astrophysique, CEA/Saclay, F-91191 Gif-sur-Yvette, France}\\
        E-mail: \texttt{lionel.prat@cea.fr}
}

\author{J. Rodriguez}{
  address={DSM/IRFU/Service d'Astrophysique, CEA/Saclay, F-91191 Gif-sur-Yvette, France}\\
        E-mail: \texttt{lionel.prat@cea.fr}
}

\begin{abstract}
The onset of most microquasar outbursts is characterized by a state transition between a Low/Hard State (LHS) and a High/Soft State (HSS). Besides drastic spectral and timing changes, this transition often shows a discrete ejection event detectable in the radio range. However, the exact nature of the ejected material and the mechanisms that give birth to these phenomena are yet to be unraveled. Recent simultaneous radio and X-ray observations on several sources point to a coronal nature of the ejected material. In the cases of GRS 1915+105, XTE J1550-564, and the 2002 outburst of GX~339-4, the flux of the Compton component decreases sharply just before an ejection is detected in the radio range. Finally, in the case of H1743-322, drastic physical changes occurred in the corona just before the state transition, compatible with the disappearance of part of this medium. Thus, the behaviour of at least 4 microquasars points in the direction of an ejection of the corona at the state transition, feature that is yet to be confirmed (or infirmed) in the case of other available sources.\end{abstract}

\maketitle


\section{Introduction}
Up to now, about 20 black-hole X-ray transients have been observed, along with $\sim$20 more candidate black-hole binaries (McClintock et al. \cite{Revue_BHB}). This led to the identification of several spectral states, depending both on the spectral and temporal characteristics of the source. The onset of a microquasar outburst usually occurs in a Low/Hard State (LHS, following the classification by Homan et al. \cite{Revue_States}), dominated by a hard X-ray emission from a relativistic jet and/or from a so-called coronal component, combined with strong temporal variability and Quasi-Periodic Oscillations (QPOs). The object then evolves, through a state transition, into a High/Soft State (HSS) dominated by soft X-ray emission from the accretion disk and less temporal variability. Another state transition usually occurs at the end of the outburst, when the source goes back in the LHS. This simple general picture is often complicated by the presence of intermediate states, mixture of the canonical LHS and HSS: the Hard InterMediate State (HIMS) and the Soft InterMediate State (SIMS).

A key to fully understand the physics of these states may reside in deeper studies of the state transitions. These phenomena involve fast and drastic changes in the system, and show strong connections between the relativistic jet and the accretion disk. The first transition, between the HIMS and the SIMS, often shows a discrete ejection event detectable in the radio range, simultaneous with a rapid increase of the disk flux and a quenching of the jet emission. Thus, the use of simultaneous radio and X-ray observations is necessary to adequately monitor the events happening during this transition. Such simultaneous monitoring is available for a handful of sources only, but their study may already lead to good constraints on the evolution of the coronal part of the system, as we show here.

\section{GRS 1915+105 and XTE J1550-564: probable ejection of the corona}
GRS 1915+105 is a very peculiar source, which shows a wider and faster range of behaviour than the other microquasars. This source has also stayed in outburst at least for the last $\sim$15 years, giving a rare luxury of data to understand accretion and ejection physics. In the X-ray range, the \rxte\ observations allowed the identification of 12 separate classes (Belloni et al. \cite{Belloni}), which could be interpreted as transitions between three basic spectral states: a Hard Intermediate one (State C) and two Soft Intermediate ones (State A and State B). A joint \inte\ and \rxte\ monitoring campaign follows the source since 2003. From these data, we have recently pinpointed the strong similarities between every spectral classes characterized by X-ray cycles (including the $\lambda$ class, Rodriguez et al. \cite{1915}). In the radio range, these states show, in particular, strong radio variability, interpreted as discrete ejections of matter.

\begin{figure}[b]
\centering
\includegraphics[width=\textwidth]{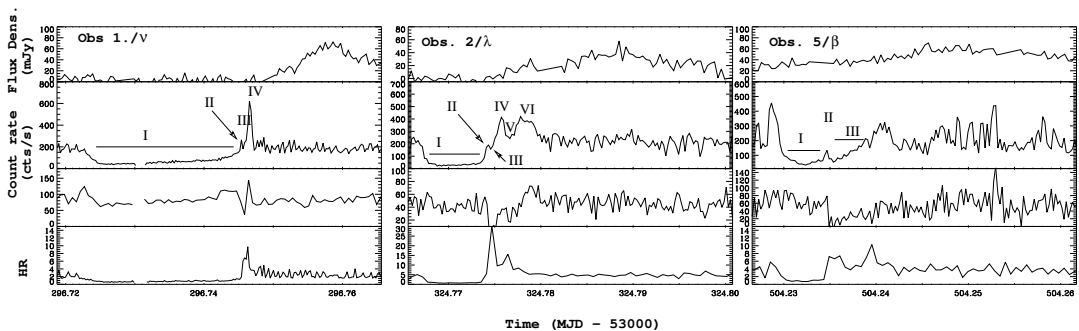}
\caption{Three cycle lightcurves, corresponding to the $\nu$, $\lambda$ and $\beta$ spectral classes of GRS 1915+105. From top to bottom, the panels represent respectively the Ryle 15GHz, the \inte/JEM-X 3-13 keV, the \inte/ISGRI 18-50 keV lighcurves, and the 3-13 keV/18-50 keV Hardness Ratio (from Rodriguez et al. \cite{1915}).}
\label{GRS1915}
\end{figure}

Fig. \ref{GRS1915} shows radio and X-ray lightcurves of GRS 1915+105 for selected spectral classes (two top panels). On these curves, X-ray dips (I) are followed by a spike (II) and, $\sim$0.3 hour after the spike, by a radio ejection. In these observations, we showed that each spike is indicative of the disappearance of the Compton component from the X-ray spectra. This points to an ejection of (part of) the corona, which is then detected in the radio range.

\medskip

A similar behaviour was already reported in XTE J1550-564 by Rodriguez et al. \cite{1550}, during its 2000 outburst. Its spectral evolution before and after the first state transition displays a significant decrease of the power-law flux, while the disk-blackbody flux remains approximately constant. Given the relative constancy of the photon index on those days, this may suggest that there is less Compton up-scattering of the soft photons, at about the same time a radio ejection is detected. It is thus tempting to consider that the Compton medium has been blown away.

\begin{figure}[t]
        \resizebox{7.7cm}{!}{\includegraphics  {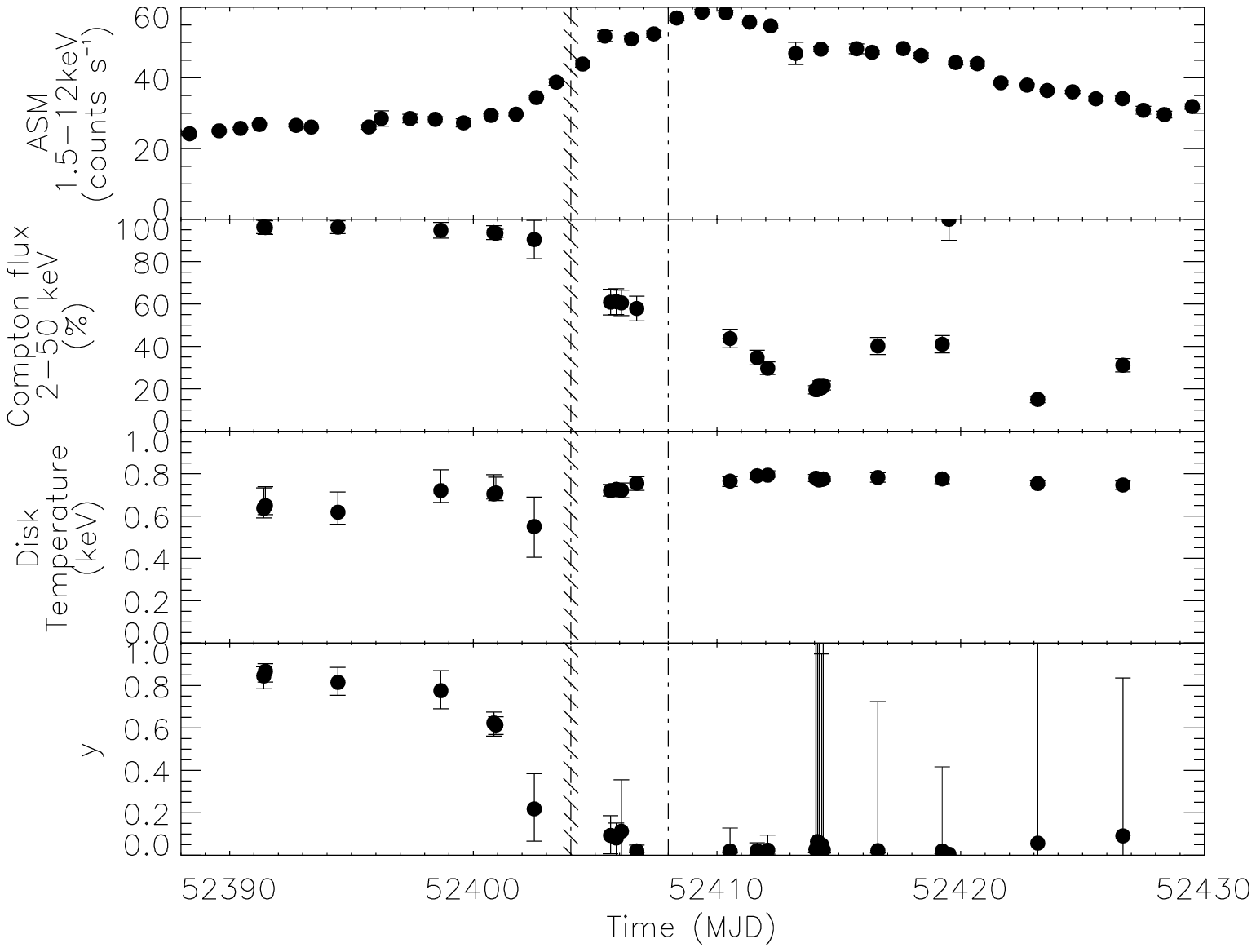}}
        \resizebox{7.7cm}{!}{\includegraphics  {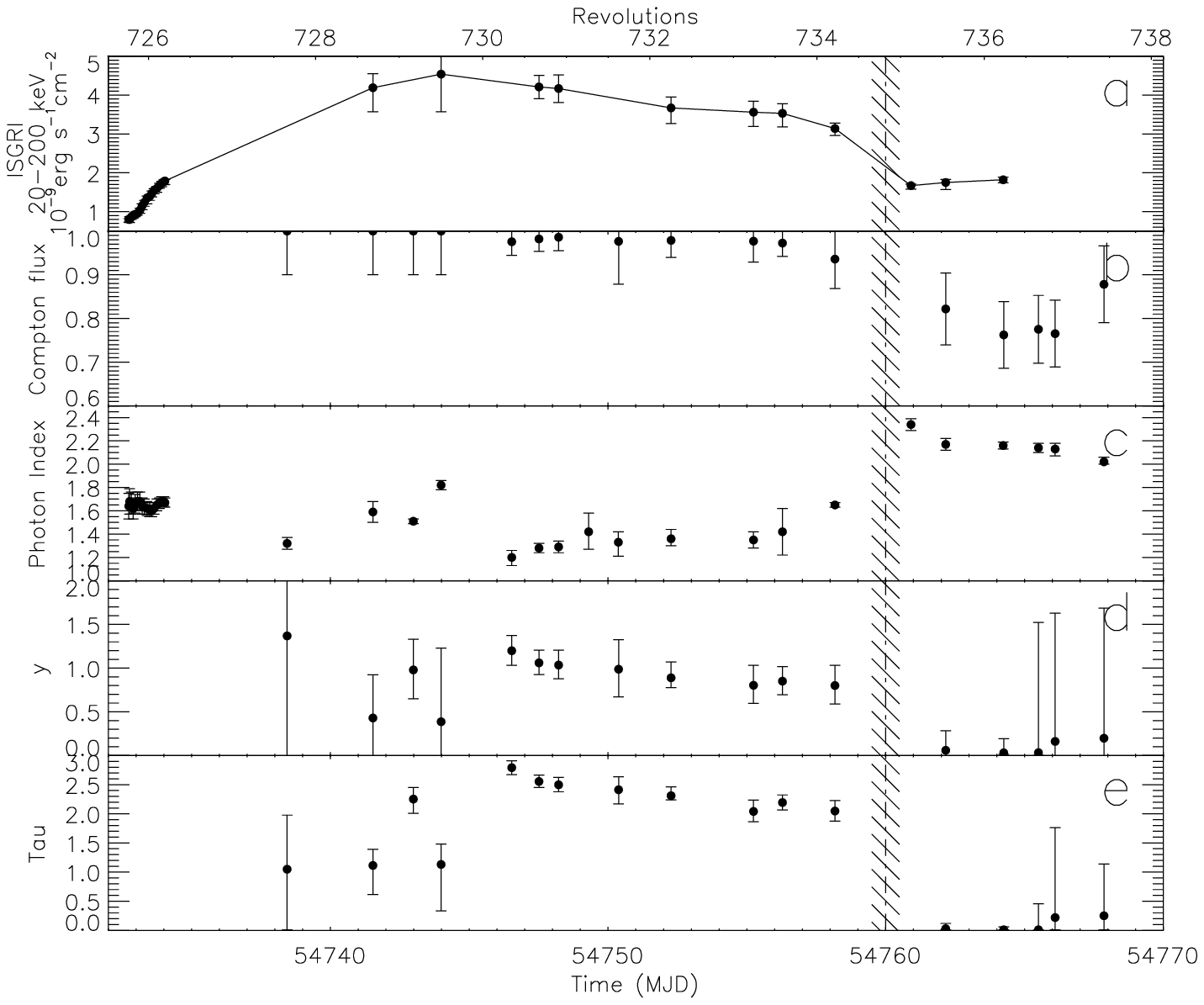}}
\caption{{\bf Left:} Spectral evolution of GX 339-4 at the beginning of its 2002 outburst. From top to bottom, the panels represent respectively the \rxte/ASM 1.5-12 keV lightcurve, the Compton component flux compared to the total flux, the disk temperature, and the Comptonization parameter $y=kT_e max(\tau, \tau^2)$. The vertical dot-dashed line marks the detection of a radio ejection, and the shaded part marks the transition from LHS to HSS. {\bf Right:} Spectral evolution of H1743-322 over its September-November 2008 outburst. From top to bottom, the panels represent respectively the \inte/ISGRI light-curve, the evolution of the Compton flux, the photon index, the $y$ parameter and the optical depth $\tau$. The shaded part marks the transition from LHS to HIMS.}
\label{GX339-4}
\end{figure}

\section{Two more candidates: GX 339-4 and H1743-322}
GX 339-4 is a well-known X-ray binary, which underwent at least 10 outbursts since its discovery in 1972 (Kong et al. \cite{Kong}). The two outbursts that occurred in 1998 and 2002 were followed by simultaneous \rxte\ observations and radio observations. The strong anticorrelation between the radio and hard X-ray fluxes in 1998 have already led Fender et al. \cite{339-4} to suspect a reduction of the coronal medium at the beginning of the HSS. At that time, only \rxte/ASM observations were available, which did not permit any precise spectral analysis of this state transition.

The 2002 outburst was followed by \rxte/PCA (e.g. Homan et al. \cite{GX339}), from which some spectral results are reported on Fig. \ref{GX339-4}, left. The evolution of the Comptonization parameter $y \propto kT_e max(\tau, \tau^2)$ (Rybicki \& Lightman \cite{yy}) and of the optical depth $\tau$ are of particular interest. Indeed, $y$ is a measure of the efficiency of the Comptonization process, while $\tau$ is proportional to the product of the density of the medium $\rho$, and its typical size R: $\tau \propto \rho R$. Thus, a diminution of $\tau$ either corresponds to a diminution of the density of the coronal medium, or to a diminution of its size. In both cases, the medium looses matter just prior to the state transition. Moreover, this evolution occurred at the time of a radio ejection, detected on MJD 52408, which is indicated by a vertical line on the plot. This shows that, again, in the case of GX 339-4, the state transition is characterized by a major change in the coronal medium, which may indicate that part of this medium is ejected by the system.

\medskip

H1743-322 is an X-ray binary discovered in 1977, which underwent a new outburst in September 2008 (Kuulkers et al. \cite{Kuulkers}). Several X-ay satellites followed this outburst almost every second day until December 2008 (Prat et al. \cite{lettre}). Using a model similar to the one used with GX 399-4, we were able to monitor the coronal evolution during this outburst. The optical depth of this medium decreased sharply just at a transition from a LHS to a HIMS (Fig. \ref{GX339-4}, right). This indicates that, for H1743-322 also, major changes occurred in the corona at the state transition. Unfortunately, no radio coverage was available at this time, and thus we could not detect any discrete ejection of matter at this time.

\section{Conclusions}
In the cases of the 4 sources studied here, a similar behaviour is detected. An increase of the soft X-ray spectrum and a softening of the X-ray flux leads to an ejection that occurs $\sim$1-3 days after. Spectral fits show that prior to this ejection, the coronal medium parameters change drastically; in two cases the corona disappears, while in the two others the product $\rho R$ shrinks, which may indicate a disappearance of part of the corona. It is thus tempting to link the coronal changes to the ejection of matter. Note that in 2 other sources, XTE J1859+226 and XTE J1748-288, a similar behaviour was seen (Rodriguez \& Prat 2008 \cite{proc}).

This interpretation still needs to be confirmed by a detailed analysis of the spectral behaviour of more back-hole binaries: the coronal emission lies in the high-energy end of the available X-ray spectra, which renders it difficult to detect and monitor. However, if this phenomenon is confirmed, it will put strong constraints on disk accretion models and, ultimately, provide a new insight on the interplay between the different components that surround an accreting compact object.


\bibliographystyle{aipproc}

\end{document}